\documentclass[aps,12pt,amssymb,preprintnumbers,amsmath,amsfonts,nofootinbib]{revtex4}
\usepackage{latexsym}
\usepackage{graphicx}

\usepackage{amsbsy}
\usepackage{amstext}
\usepackage{amssymb}
\usepackage{amsmath}

\newcommand{\be}{\begin{equation}}
\newcommand{\ee}{\end{equation}}
\newcommand{\ba}{\begin{eqnarray}}
\newcommand{\ea}{\end{eqnarray}}
\newcommand{\nn}{\nonumber}
\newcommand{\beq}{\begin{equation}}
\newcommand{\eeq}{\end{equation}}

\newcommand{\Eq}[1]{Eq.~\eqref{#1}}
\newcommand{\vL}{{(v \cdot L)}}
\newcommand{\lnLL}{\ln\frac{l_0 + l}{l_0 - l}}
\usepackage{slashed}
\usepackage{hyperref}

\date{\today}

\begin{document}

\title{Power corrections to the HTL effective Lagrangian of QED}

\author{Stefano Carignano}
\email{carignano@lngs.infn.it}
\affiliation{INFN, Laboratori Nazionali del Gran Sasso, Via G. Acitelli 22, I-67100 Assergi (AQ), Italy
}
\author{Cristina Manuel}
\email{cmanuel@ice.csic.es}
\affiliation{Instituto de Ciencias del Espacio (ICE, CSIC) \\
C. Can Magrans s.n., 08193 Cerdanyola del Vall\`es, Catalonia, Spain
and \\
 Institut d'Estudis Espacials de Catalunya (IEEC) \\
 C. Gran Capit\`a 2-4, Ed. Nexus, 08034 Barcelona, Spain
}
\author{Joan Soto}
\email{joan.soto@ub.edu}
\affiliation{Departament de F\'\i sica Qu\`antica i Astrof\'\i sica 
                   and Institut de Ci\`encies del Cosmos,
        Universitat de Barcelona,
        Mart\'\i $\,$ i Franqu\`es 1, 08028 Barcelona, Catalonia, Spain.}


\begin{abstract}
We present compact  expressions for the power corrections to the hard thermal loop (HTL) Lagrangian of
QED in $d$ space dimensions. These are corrections of order $(L/T)^2$, valid for momenta $L \ll T$, where $T$ is the temperature. 
In the limit $d\to 3$ we achieve a consistent regularization of both infrared and ultraviolet divergences, which respects the gauge symmetry of the theory. Dimensional regularization also allows us to witness subtle cancellations
of infrared divergences.
 We also discuss how to generalise our results in the presence of a chemical potential, so as to obtain the power corrections to the hard dense loop (HDL) Lagrangian.
\end{abstract}

\maketitle

\section{Introduction}

At high temperature $T$ the physics of the so called soft scales, or scales or order $eT$, where $e$ is the QED gauge coupling constant, is properly described by
the hard thermal loop (HTL) effective field theory (EFT)  \cite{Pisarski:1988vd,Braaten:1989mz,Frenkel:1989br}.
The HTL effective Lagrangian of QED reads \cite{Braaten:1991gm,Braaten:1992jj,Pisarski:1997cp}

\begin{equation}
{\cal L}^{(1)}_{\rm HTL} = \frac{e^2 }{2} \int \frac{d^3 q}{(2 \pi)^3} \Bigg\{ \frac{ 2 n_F(q) }{q} \left(F_{\rho \alpha}   \frac{v^\alpha v^\beta}{(v \cdot \partial)^2 } F^{\rho}_{\,\,\beta}  \right) -
\frac{ 2(n_F(q) + n_B(q)) }{q}  
\left( {\bar \psi} \frac{\slashed{v}}{(iv \cdot D)}\,\psi  \right) \Bigg\}
\ ,
\end{equation} 
where $v^\mu = q^\mu/|{\bf q}|$ is a light-like vector, and $n_F(x) = (e^{x/T} + 1)^{-1}$ and $n_B(x) = (e^{x/T} - 1)^{-1}$ stand for the fermionic/bosonic  thermal distribution functions, respectively.

Corrections to the HTL results for a number of physical observables have been discussed since long, mostly for the static case in the imaginary time formalism (see \cite{Kraemmer:2003gd} and references therein).
However, little is known on the structure of higher order corrections to the HTL Lagrangian itself. This is exceptional in the EFT realm, since for most of them one can write down the corresponding terms of the Lagrangian 
at any desired order in the expansion parameter (see for instance \cite{Bijnens:2006zp,Brambilla:2004jw}). Each term consists of an operator and a matching coefficient. The set of operators that enter at a given order is fixed by dimensional analysis and the symmetries of
the fundamental theory. Fixing the matching coefficient requires either a calculation in the fundamental theory or a comparison with  experiment. In contradistinction, for the HTL Lagrangian, it is not even known the kind 
of terms that would appear at next-to-leading order (NLO). The problem is related to the fact that the HTL Lagrangian at leading order (LO) is 
non-local, and it is not apparent which kind of non-localities will be generated at higher orders. In this paper, we shed light on this issue. We focus on the power 
corrections to the HTL Lagrangian, namely corrections of the type $L/T$, where $L$ is a momentum scale such that $L  \ll T$. We consider only contributions generated in one-loop diagrams, which  have been  also called 
one-loop hard corrections in Ref.~\cite{Mirza:2013ula}. 
These are expected to be  the leading corrections  when the momentum of the soft modes $L$ lies between $eT$ and $T$, but as we will discuss, they turn out to be of the same relevance
as other perturbative corrections ~\cite{Mirza:2013ula}.

In Ref.~\cite{Manuel:2016wqs} the leading power correction to the HTL photon polarisation tensor was computed using the on-shell effective field theory (OSEFT) \cite{Manuel:2014dza}.
 We show here that it is possible to write in a simple compact form a term in the effective Lagrangian that reproduces it. 
The expression in $d$ spatial dimensions reads

\begin{equation}
\label{npc-HTL}
{\cal L}^{(3)\gamma}_{\rm HTL} = \frac{e^2 \nu^{3-d}}{4} \int \frac{d^d q}{(2 \pi)^d}  \frac{ 1- 2 n_F(q) }{q^3} \Bigg \{ F_{\rho \alpha}   \frac{v^\alpha v^\beta}{(v \cdot \partial)^4 } \partial^4 F _{\beta}^{\,\,\rho}  \Bigg \} \,,
\end{equation} 
where $\nu$ is the renormalization scale, and now $v^\mu$ is a light-like vector in $d+1$ dimensions. 

We also calculate here the leading power correction to the fermionic sector in the HTL Lagrangian, which can be expressed as
\ba 
{\cal L}^{(3)\,\psi}_{\rm HTL} &=&  \frac{e^2 \nu^{3-d}}{4}   \left(d-1  \right) \left[ \int \frac{d^d q}{(2 \pi)^d}  \frac{  n_F(q) + n_B(q) }{q^3}  \Bigg \{ 
\, {\bar \psi}\,D^2\frac{\slashed{v}}{(iv \cdot D)^3}D^2\,\psi  \Bigg \}\right.\nn\\
& + & \left.  \int \frac{d^d q}{(2 \pi)^d}  \frac{ 1+2 n_B(q) }{2 q^3} \,  \Bigg \{ {\bar \psi}\left(D^2i\slashed{D}\frac{1}{(iv \cdot D)^2}+ \frac{1}{(iv \cdot D)^2}i\slashed{D} D^2\right)\psi  \Bigg \}\right]  +{\cal O} (e^3) \ .
\label{nhtlf}
\ea

In the remaining part of the Letter we briefly explain how to obtain the terms in the effective Lagrangian displayed above. We will also 
comment on how our results can be generalized in the presence of a chemical potential $\mu$.
We use natural units $\hbar= c = k_B =1$, metric conventions $g^{\mu \nu} = (1, -1,-1,-1)$ and use capital letters to denote 4-momenta, so
$P^\mu = (p_0, {\bf p})$ and $p = |{\bf p}|$, so that $P^2 = p_0^2 -p^2$.

\section{Power corrections to the HTL Photon self-energy}

In QED the retarded  photon polarization
tensor in the Keldysh representation  of the real time formalism  (RTF)  reads  \cite{Carrington:1997sq} 
\begin{equation}
\Pi^{\mu\nu}_R(L) =-\frac{ie^{2}}{2}\int\frac{d^4q}{(2\pi)^{4}}\Big ( {\rm Tr}[\gamma^{\mu}S_{S}(K)\gamma^{\nu}S_{R}(Q)]+{\rm Tr}[\gamma^{\mu}S_{A}(K)\gamma^{\nu}S_{S}(Q)]\Big) \ ,
\label{eq:IntDefFull} 
\end{equation}
where  $K= Q- L$, and the fermion retarded/advanced and symmetric propagators 
are
\begin{equation}
S_{R/A}(Q)=\frac{\slashed{Q}}{Q^{2} \pm i\textrm{sgn}(q_{0})\eta} \;, \qquad
 S_{S}(Q)=-2\pi i \slashed{Q}\,(1-2n_{F}(|q_{0}|))\delta(Q^{2})\, ,\label{eq:propfull}
\end{equation}
respectively, with $\eta \rightarrow 0^+$.

If one makes the change of variables $Q \rightarrow - K$  in the first term of Eq.~(\ref{eq:IntDefFull}) and carries out the $q_0$ integral, then one gets
\begin{eqnarray}
\Pi^{\mu\nu}_R(L) &= & e^2 \int\frac{d^{3}q}{(2\pi)^{3}} \frac{ 1- 2 n_F(q) }{q} \left( \frac {   2 q v^\mu v^\nu - (v^\mu L^\nu + v^\nu L^\mu) + g^{\mu \nu} \, v \cdot L }{ v \cdot L- \frac{L^2}{2q } + i \,\textrm{sgn}(q -l_{0})\eta } \right. 
\nonumber \\ 
&-& \left. \frac{ 2 q {\tilde v}^\mu {\tilde v}^\nu - ({\tilde v}^\mu L^\nu + {\tilde v}^\nu L^\mu) + g^{\mu \nu} \, {\tilde v} \cdot L }{ {\tilde v} \cdot L +\frac{L^2}{2q } + i \,\textrm{sgn}(q +l_{0})\eta  }  \right) \ ,
\label{totalPi}
\end{eqnarray}
where $v^\mu = (1 , {\bf q}/q)$ and ${\tilde v}^\mu = (1, -{\bf q}/q)$. The first and second terms in Eq.~(\ref{totalPi}) correspond to the particle and antiparticle contributions, respectively.
As we are assuming a thermal bath where parity is conserved, one can carry out the change of variables ${\tilde v}^\mu \rightarrow v^\mu$ in the piece that describes the antiparticle contribution.
We consider the situation where the external momentum is soft, $L \ll Q$, so that the integrand of Eq.~(\ref{totalPi}) can be expanded in $L^2/2 q^2$. The expansion contains both UV and IR divergent terms,
  and a regularisation must be introduced. We use dimensional regularisation (DR) to treat the divergences of both the vacuum and thermal
part of the integrals. More specifically, 
\begin{eqnarray}
\Pi^{\mu\nu}_{(1)}(L) &= & 2e^2\nu^{3-d} \int\frac{d^{d}q}{(2\pi)^{d}} \frac{ 1- 2 n_F(q) }{q} \left( \frac {  v^\mu v^\nu L^2}{ (v \cdot L)^2} - \frac{v^\mu L^\nu + v^\nu L^\mu}{ v \cdot L}+ g^{\mu \nu} \right) \ ,
\label{PiHTL}
\\ 
\Pi^{\mu\nu}_{(3)}(L) &= & 2 e^2 \nu^{3-d} \int\frac{d^{d}q}{(2\pi)^{d}} \frac{ 1- 2 n_F(q) }{q^3} \frac{L^4}{ 4 (v \cdot L)^2} \left( \frac {  v^\mu v^\nu L^2}{ (v \cdot L)^2} - \frac{v^\mu L^\nu + v^\nu L^\mu}{ v \cdot L}+ g^{\mu \nu} \right) \ ,
\label{powerPiHTL}
\end{eqnarray}
where retarded boundary conditions are taken into account with the prescription $l_0~\rightarrow~l_0 + i 0^+$.
After functional differentiation, 
it is now easy to check that the polarization tensor (\ref{powerPiHTL}) is obtained from the effective Lagrangian Eq.~(\ref{npc-HTL}). 
We also note that  higher-order power corrections can easily be inferred by further expanding  Eq.~(\ref{totalPi}). It is not difficult to realize that
higher-order corrections are extracted by multiplying the integrand of Eq.(\ref{PiHTL}) by
\begin{equation}
  \left( \frac{L^4}{4 q^2 (v \cdot L)^2} \right)^n \ , \qquad n=1,2, 3, \,\ldots
 \end{equation}
providing the form of the corresponding higher-order terms in the effective Lagrangian.

Eq.~(\ref{PiHTL}) is finite when $d\to 3$ and corresponds to the HTL polarisation tensor. The longitudinal and transverse components of the HTL photon self-energy read
\begin{eqnarray}
\Pi^{L}_{(1)} &= & -\frac{m_D^2}{2}  \left(1-\frac{l_0}{2l}\lnLL \right) \ ,\\
\Pi^{T}_{(1)} &= & -\frac{m_D^2}{2}\left[1+ \frac{L^2}{l^2} \left(1-\frac{l_0}{2l}\lnLL\right ) \right] \ ,
\end{eqnarray}
respectively, where $m_D^2 = e^2 T^2 /3$ is the Debye mass.

When $d\to 3$, Eq.~(\ref{powerPiHTL}) is UV divergent only, as the IR divergence of the vacuum part is exactly cancelled by the IR divergence of the
thermal contribution to the integral. It is  important to use the same IR regulator for the vacuum and thermal parts to effectively achieve this cancellation.
 A proper evaluation of the DR regulated expressions (see Appendix \ref{DR-formulas}) shows that Eq. (\ref{powerPiHTL}) reproduces exactly the power correction
computed with the OSEFT in Ref.~\cite{Manuel:2016wqs}, namely for $d=3+2\epsilon$ one gets 

\begin{eqnarray}
\label{final-Long}
\Pi^{L}_{(3)} &= &\frac{\alpha}{ 3 \pi }\left[ \frac{l^2}{ \epsilon} +  2 l^2
 \left( \ln \frac{\sqrt{\pi}Te^{- \gamma_E/2}}{2 \nu}  -1 \right) +
\left(2l^2-L^2\right)\left(1- \frac{l_{0}}{2 l} 
 \,{\rm ln\,}{\frac{l_0+ l}{l_0-{ l}}} \right) \right] \,, \label{final-L}
\\
 \Pi^{T}_{(3)}  &= & \frac{ 2\alpha L^2}{3\pi } \left[ \frac{1}{2 \epsilon} +  \left( \ln \frac{\sqrt{\pi}T e^{- \gamma_E/2}}{2 \nu}  -1 \right) 
+  \frac 14 + \left( 1+\frac{L^2}{4 l^2}\right) \left(1-\frac{l_0}{2 l}
 \,{\rm ln\,} \frac{l_0+l}{l_0-l}  \right)  \right] \,,
\label{final-T}
\end{eqnarray}
where $\alpha$ is the electromagnetic fine structure constant.
The divergent pieces above are eliminated with the QED counterterm that takes into account the photon wavefunction renormalization 
\footnote{Our results differ from those displayed in the Appendix of Ref.\cite{Weldon:1982aq}. There the vacuum part was regulated with
DR, while  to compute the thermal part  a different regulator was introduced. That regulator treats the radial part of the divergent integrals as in DR, but the angular integrals are
computed in $d=3$ dimensions. Our explicit computations show that when this is done, the Ward identity is violated \cite{Manuel:2016wqs}.}.

Although so far we have focused our analysis on systems at very high $T$, our results are easy to generalize in the presence of a chemical potential $\mu$. This requires
 taking into account that fermion and anti-fermion degrees of freedom have
different distribution functions in the Keldysh symmetric propagators \cite{Mallik:2009pj}. For the photon self-energy, 
all our results remain valid after  simply replacing 
\be
\label{gen-withmu}
n_F(q) \rightarrow \frac 12 \Big[ n_F( q -\mu) + n_F (q + \mu) \Big] \ 
\ee
 in Eqs.~(\ref{npc-HTL}),~(\ref{PiHTL}) and (\ref{powerPiHTL}). After performing the radial integrals (see Appendix A for explicit expressions at finite $T$ and $\mu$), we can write general expressions for any value
 of $\mu$. In particular,  the result for vanishing temperature and nonzero chemical potential
can be straightforwardly obtained from \Eq{final-T} by the replacement
 \be
 \ln \frac{\sqrt{\pi} e^{-\gamma^E/2} T }{2\nu} \;\; \rightarrow \; \; \ln \frac{ e^{\gamma^E/2} \mu }{\sqrt{\pi}\nu} \,.
 \ee
 
 Thus, with this replacement, we also obtain the power corrections to the hard dense loops (HDL) 
 \cite{Manuel:1995td}.

\section{Power Corrections to the HTL Fermion Self-Energy} 

\noindent
The retarded fermion self-energy in the Keldysh representation  of the RTF  reads
\begin{equation}
\Sigma_R(L) = \frac{ i e^2}{2} \int\frac{d^4q}{(2\pi)^{4}}\Big ( \gamma_{\mu}S_{S}(Q)\gamma_{\nu} D^{\mu \nu}_A(K) + \gamma_{\mu}S_{R}(Q)\gamma_{\nu}D^{\mu \nu}_{S}(K) \Big) \ ,
\label{eq:SigmaR}
\end{equation}
with   $K= Q- L$. 
In a covariant gauge the photon propagator is given by
\begin{equation}
D^{\mu\nu}_i (Q) = - \left( g_{\mu \nu} + \xi \frac{Q^\mu Q^\nu}{Q^2} \right) \Delta_i (Q) \ , \qquad  i= R, A, S
\label{eq:Dmunu}
\end{equation}
where $\xi$ encodes the gauge-fixing parameter dependence and 
\begin{equation}
\Delta_{R/A}(Q)=\frac{1}{Q^{2} \pm i\textrm{sgn}(q_{0})\eta} \ , \qquad
 \Delta_{S}(Q)=-2\pi i \,(1+2n_{B}(|q_{0}|))\delta(Q^{2})\, ,\label{eq:propfull}
\end{equation}
are the retarded/advanced and symmetric bosonic propagators, respectively.
Let us focus first on the pieces that do not depend on $\xi$,
 and postpone the discussion of the others to the next Subsection.

To proceed with the evaluation of the fermion self-energy we carry out similar manipulations as those we performed in the previous Section,
 which require  parity to be a good symmetry of the system. 
We then expand the integrands of Eq.~(\ref{eq:SigmaR}), assuming 
$L \ll Q$, and arrive at
\begin{eqnarray}
\label{fermiHTL}
\Sigma^{\xi=0}_{(1)}& = & - \frac {e^2\nu^{3-d}}{2} (1 -d)\, \int\frac{d^{d}q}{(2\pi)^{d}} \frac{  n_F(q) + n_B(q) }{q}  \frac{\slashed{ v}} { v \cdot L} \ ,\\
\label{powerfermiHTL}
\Sigma^{\xi=0}_{(3)} & = & -  \frac {e^2\nu^{3-d}}{2} (1 -d)\,\int\frac{d^{d}q}{(2\pi)^{d}} \left( \frac{  n_F(q) + n_B(q) }{2 q^3}  \frac{ L^4 \,\slashed{ v}} { (v \cdot L)^3}  
- \frac{ 1 + 2 n_B(q)}{2 q^3} \frac { L^2 \slashed{L}}{(v \cdot L)^2} \right) \,.
\end{eqnarray}
Eq.~(\ref{fermiHTL}) describes the HTL fermion self-energy, while Eq.~(\ref{powerfermiHTL}) describes its first power correction.
Now it is easy to see that from the effective Lagrangian (\ref{nhtlf}) one can obtain the HTL power correction (\ref{powerfermiHTL}). Note that we wrote Eq.~(\ref{nhtlf}) in a way that  is manifestly gauge invariant and
Hermitian, replacing  derivatives by covariant derivatives. 
 We chose a particular ordering of the differential operators in the Lagrangian. Any alternative ordering agrees with ours up to terms suppressed by factors of $e$, but this can only be fixed by computing 
higher-point Green functions.

While Eq.~(\ref{fermiHTL}) is finite when $d\to 3$, and can be easily evaluated, Eq.~(\ref{powerfermiHTL}) contains both UV and IR divergencies.  Only IR divergencies appear in the first term 
of Eq.~(\ref{powerfermiHTL}) if one evaluates separately the fermionic and the bosonic contributions, but the divergencies  cancel in the sum
(see Eqs.~(\ref{DR-bos}) and  (\ref{DR-fer})), so that in the limit $\epsilon \to 0$  the integral associated with the first term remains finite. Since DR sets all dimensionless integrals to zero, 
the singularity appearing in the second term of Eq.~(\ref{powerfermiHTL}) actually corresponds to
an UV divergence, that may be absorbed by the fermion wave-function renormalization. Note also that the linear IR divergences induced by the bosonic distribution function are set to zero in DR.
More explicitly, we arrive at

\ba
\label{HTLf}
\Sigma^{\xi=0}_{(1)} & = & m_f^2\left[\frac{\gamma^0}{2l}\lnLL+\frac{\boldsymbol{\gamma}\cdot\mathbf{l}}{l^2}\left(1-\frac{l_0}{2l}\lnLL \right)\right] \ , \\
\Sigma^{\xi=0}_{(3)} & = & \frac{\alpha}{4\pi}  \slashed{L} \left[  \frac{1}{\epsilon} - \gamma_E + 1 + \ln\left(\pi \frac{T^2}{\nu^2}\right) - \frac{l^0}{l} \lnLL \right] \ ,
\label{NHTLf}
\ea
where  $m_f^2 = e^2 T^2 /8 $ is the electron thermal mass.

As in the case for the photon self-energy, in the presence of a chemical potential the same Eqs.~(\ref{nhtlf}), (\ref{fermiHTL}) and (\ref{powerfermiHTL}) remain valid 
if we replace the fermionic distribution function as prescribed in Eq.~(\ref{gen-withmu}).

\subsection{Gauge-fixing dependent pieces in the fermion self-energy}

These contributions come from the  $\xi$-dependent pieces in the photon propagator 
in \Eq{eq:SigmaR}. Here a regularization of the $\delta(Q^2)/Q^2$ term which arises in the photon propagator must be given. We follow the prescription in \cite{Weldon:1982bn},
substituting 
 $\delta(Q^2)/Q^2 \rightarrow -d/dq_0^2 \delta(Q^2)$. After integrating by parts,  performing the same sort of operations that we did in the previous cases, and
 expanding for  $L \ll Q$, we find that the first non-vanishing contribution is

\begin{align}
\Sigma^{\xi}_{(3)} & =  \xi \frac {e^2\nu^{3-d}}{4} L^2 \int\frac{d^{d}q}{(2\pi)^{d}} \Bigg\lbrace \frac{  1+ 2n_B }{2q^3}  \frac{\gamma^0}{\vL} + \left[ \frac{1}{q^2} \frac{dn_B}{dq} - \left(\frac{1+2n_B}{2q^3}\right) \right]  \frac{\slashed{v}}{\vL} \nonumber\\
& - \left[ \frac{3 - 4 n_F + 2n_B}{2q^3}\right] \frac{\slashed{L}}{\vL^2} - \frac{1+2n_B}{2q^3} \frac{l_0 \slashed{v}}{\vL^2} +  \frac{1 - n_F + n_B}{q^3} \frac{L^2 \slashed{v}}{\vL^3} \Bigg\rbrace \ .
\end{align}
After carrying out radial and angular integrals in DR , we arrive at
\beq
\Sigma^{\xi}_{(3)} =  \frac{\alpha \xi}{4\pi} \left[ \slashed{L} \left( \frac{1}{\epsilon} + 1 + \ln\left(\pi \frac{T^2 e^{-\gamma_E}}{\nu^2}\right) - \frac{l^0}{l} \lnLL \right) - L^2 
\frac{\boldsymbol{\gamma}\cdot \mathbf{l}}{l^2} \left(1-\frac{l^0}{2l} \lnLL \right) \right]\,.
\label{nhtlfxi}
\eeq
Once again,  the $1/\epsilon$ pole above is of UV nature and can be absorbed by a wave function renormalization. Moreover, we can see that the first piece in Eq.~(\ref{nhtlfxi})
 is remarkably similar to the $\xi=0$ contribution to the fermion self-energy, Eq.~(\ref{NHTLf}), so the two cancel out if $\xi=-1$. Our results are thus consistent with the known fact that
 in the Feynman gauge, $\xi=-1$,  there is no fermion wave function
renormalization. 

Formally,  the non-local pieces in Eq.(\ref{nhtlfxi}) can  be removed by carrying out the following field redefinitions in the Dirac Lagrangian

\be
\delta \psi=\frac{\alpha\xi}{8\pi}\int \frac{d\Omega}{4\pi}\left(2D_0\frac{1}{v \cdot D}+ \slashed{D}\frac{\boldsymbol{\gamma}. \mathbf{v}}{v \cdot D} \right)\psi \, ,
\label{frd}
\ee
where $d\Omega$ is the solid angle in $d=3$ . However, when plugged into the HTL Lagrangian, 
this induces new ${\cal O}\, (e^4)$ contributions that should cancel out against $\xi$-dependent terms arising from the two-loop hard correction to the HTL effective action, which are of the same order \cite{Mirza:2013ula}.
In fact, similar field redefinitions allow to trade the gauge-independent piece  of the fermion self-energy Eq.~(\ref{NHTLf}) for ${\cal O} (e^4)$ terms, since it is also proportional to $\slashed{L}$. 
Actually, this also happens for the photon contributions to the power corrected Lagrangians, as we show in Appendix \ref{NHTLph}.

\section{Discussion}

We have provided compact  expressions for the power corrections to the HTL  Lagrangians in QED.
These are expected to be part of the leading corrections for momenta above the soft scale, and such that  $eT < L \ll T$, although
we have seen that they are of the same order as those arising from perturbative corrections. 
While we were mainly interested in the high temperature limit of QED, we have also explained how to obtain this type of corrections
in the presence of a chemical potential, and thus we can also get the power corrections for the HDL effective Lagrangian.

We have used DR in order to regulate 
both UV and IR divergences \cite{Manohar:1997qy}.
 In fact, we presented our results for the power corrections to the HTL Lagrangian in arbitrary space dimensions $d$.
Since DR sets power-like divergences to zero, our results turn out to be IR finite both for the photon and fermion sectors, while
the UV divergences we find can be removed by the standard counterterms in QED. DR also guaranties that gauge invariance is kept in the regulated theory, 
and thus differs from similar approaches carried out in the literature, where a cutoff, which breaks the gauge symmetry, is introduced to deal with the IR divergencies~\cite{Mitra:1999wz,Wang:2004tg}.

It would be interesting to compute the two-loop hard corrections to the photon and fermion self-energies, as these would allow us to obtain the remaining NLO terms to the HTL Lagrangian.
While those computations might be hard, we expect that the effective field theory methods developed in Ref.\cite{Manuel:2016wqs}
could pave the way. They should be carried out using DR in order to be matched consistently with the power corrections presented here, particularly in order to check the cancellation of the gauge parameter dependence, which was reshuffled into these contributions by the field redefinitions in (\ref{frd}).

In order to extract 
corrections to many physical properties of electromagnetic plasmas, such as damping rates and transport coefficients, we not only need the NLO terms in the HTL Lagrangian (the power corrections calculated here
 plus the two-loop hard corrections mentioned above), but also loop calculations with the LO HTL Lagrangian, which should also be carried out in DR for consistency. In some cases, due to Bose enhancement,  
they give the
 leading correction, for instance, to the thermal fermion mass and damping rate~\cite{Mirza:2013ula,Carrington:2008dw}.

\acknowledgements

We have been supported by the MINECO (Spain) under the projects  FPA2013-43425-P and FPA2016-81114-P.
This work was also supported by the COST Action CA15213 THOR. J.S. has also been supported by the projects FPA2016-76005-C2-1-P projects (Spain),
and the 2014-SGR-104 grant (Catalonia).

\appendix

\section{Dimensional regularisation: some useful formulas}
\label{DR-formulas}

The momentum integral in arbitrary $d$ spatial dimensions is given by \cite{Laine:2016hma}

\beq
\int \frac{d^d q}{(2\pi)^d} = \int_0^\infty dq \, q^{d-1} \int \frac{d\Omega_d}{(2\pi)^d} \rightarrow 
\frac{ 4 }{(4\pi)^{\frac{d+1}{2}} \Gamma( \frac{d-1}{2})} \,\int_0^\infty dq \, q^{d-1} \int_{-1}^1 d\cos\theta \sin^{d-3}\theta \ ,
\eeq
where $\theta$ parametrises an angle with respect to an external vector, and $\Gamma(z)$ stand for the Gamma function.

Most of the integrals appearing in this manuscript  are carried out in $d=3 + 2 \epsilon$ dimensions.  We collect here the relevant formulas needed to compute
the first power corrections to the HTLs.

The relevant radial integrals are 
\begin{align}
\nu^{- 2 \epsilon} \int_0^\infty dq q^{-1+2\epsilon} \,n_B(q &)=  \left(\frac{\nu}{T}\right)^{-2\epsilon} 
\Gamma (2\epsilon ) {\rm Li}_{2\epsilon}(1)
 = 
-\frac{1}{4\epsilon} -\frac{1}{2} \ln \left(\frac{2\pi T e^{-\gamma_E}}{\nu}\right) + {\cal{O}}(\epsilon) \ ,
\label{DR-bos}
\\ 
\nu^{- 2 \epsilon} \int_0^\infty dq q^{-1+2\epsilon} \,n_F (q)  & = 
\left(\frac{\nu}{T}\right)^{-2\epsilon} (1-2^{1-2\epsilon})\Gamma (2\epsilon) \zeta (2\epsilon)
= 
\frac{1}{4\epsilon} + \frac{1}{2}\ln \left(\frac{\pi Te^{-\gamma_E}}{2 \nu}\right)  + {\cal{O}}(\epsilon) \ ,
\label{DR-fer}
\\
\nu^{- 2 \epsilon} \int_0^\infty dq q^{2\epsilon} \,\frac{dn_B}{dq} & =\frac{1}{2} + {\cal{O}}(\epsilon) \ ,
\end{align}
where $\nu$ is the renormalisation scale, $\gamma_E$ is the Euler's constant, $\zeta(z)$ stands for the Riemann zeta function, and ${\rm Li}_{n}$ is the Euler polylogarithmic function
of order $n$.

At nonvanishing chemical potential, the necessary radial integral to be evaluated instead of Eq.~(\ref{DR-fer}) is
\beq
\label{Dr-fer-mu}
\frac{\nu^{- 2 \epsilon}}{2} \int_0^\infty dq q^{-1+2\epsilon} \Big[n_F (q- \mu) + n_F(q + \mu) \Big] =  \frac{1}{4\epsilon} + \frac{1}{2}\ln \left(\frac{\pi T }{2 \nu}\right)  - \frac 14 \Psi(T,\mu) + {\cal{O}}(\epsilon) \ ,
\eeq
where we defined
\beq
\Psi(T,\mu)  \equiv \psi^{(0)} ( 1 -  i \frac { \mu}{2 \pi T}) + \psi^{(0)} ( 1 + i \frac { \mu}{2 \pi T}) - 2 \psi^{(0)} ( 1 - i \frac { \mu}{ \pi T}) - 2 \psi^{(0)} ( 1 + i \frac { \mu}{ \pi T}) 
\eeq
and $\psi^{(0)} ( z)$ is the logarithmic derivative of the Gamma function.
We can consider two different limits of this expression:
at low temperatures $T \ll \mu$
 \beq
\Psi(T,\mu)  \rightarrow - 2 \ln \left ( \frac{2 \mu}{\pi T} \right) + \frac{\pi^2 }{3} \frac{ T^2}{ \mu^2} + {\cal O} \left(\frac{T^4}{\mu^4} \right) \,,
\eeq
so that when plugged into Eq.~(\ref{Dr-fer-mu}), the diverging logarithmic dependence on the $T$ disappears, as it should be the case.
For high temperature and low chemical potential instead one finds
 \beq
\Psi(T,\mu)  \rightarrow  2 \gamma_E - \frac{7 \zeta(3)}{2 \pi^2}  \frac{ \mu^2}{ T^2} + {\cal O} \left(\frac{\mu^4}{T^4} \right) \ .
\eeq
The only nontrivial angular integral we need to compute explicitly is 
\begin{align}
I_1 & \equiv  \int_{-1}^1 d\cos\theta \left(\sin^2\theta\right)^\epsilon \frac{1}{\vL} =  \int_{-1}^1 d\cos\theta \left(1-\cos^2\theta\right)^\epsilon \frac{1}{l^0 - l\cos\theta} \nonumber\\
      &  = \frac{1}{l} \left\{ \lnLL + \epsilon \left[ \ln (4) \lnLL + \text{Li}_2 \left(-\frac{2 l}{l^0 - l} \right)- \text{Li}_2\left(\frac{2 l}{l^0 + l}\right) \right] \right\} +  {\cal{O}}(\epsilon^2) \ ,
\end{align}
which obviously reduces to the standard result in $d=3$  \cite{Laine:2016hma} when $\epsilon \to 0$.

All other higher order $I_n = \int \frac{1}{\vL^n} $ can be straightforwardly obtained by differentiating $I_1$ with respect to $l^0$, while for terms containing $v_i$ at the numerator we can write  
\beq
\int  \frac{d\Omega_d}{(2\pi)^d} \frac{v^i}{\vL^n} = \frac{ 4 }{(4\pi)^{2+\epsilon} \Gamma(1+\epsilon)} \frac{l^i}{l^2} \left( l^0 I_{n} - I_{n-1} \right) \ .
\eeq

\section{Field redefinitions in the photon sector}
\label{NHTLph}

Exactly as it occurs for the fermionic Lagrangian, the power-corrected contributions to the HTL Lagrangian for the photon sector can be removed
by field redefinitions. Let us define the gauge field transformations
\be
\delta A_0=C_0 \,\int \frac{d\Omega}{4\pi} \ \frac 12 \frac{F_{0\nu} (v^\nu - {\tilde v}^\mu)}{v\cdot\partial} \quad , \quad \delta A_i= C_T \,\int \frac{d\Omega}{4\pi}\, \frac 12 \frac{F_{i\nu} ( v^\nu - {\tilde v^\nu})}{v \cdot\partial}\,,
\label{fr}
\ee
where $C_0$ and $C_T$ may contain derivative operators local in time and do not depend on $v$ and ${\tilde v}$. The transformations above induce the following contribution  to the transverse self-energy
\be
\delta \Pi^{T}_{(1)} (L)= C_T \,L^2\,\int \frac{d\Omega}{4\pi}\frac{ (v - {\tilde v}) \cdot L}{2 (v \cdot L)}= C_T \, L^2\, \left(1-\frac{l_0}{2l}\lnLL\right) \,.
\ee
Then by taking
\be
C_T =-\frac{2\alpha}{3\pi}\left( 1+\frac{L^2}{4 l^2}\right)\,,
\ee
we can remove the last term in (\ref{final-T}). Note that the operator $C_T$ is non-local in space but local in time ( $L^2 \rightarrow \partial^2$, and $l^2 \rightarrow -\boldsymbol{\partial}^2$ after a Fourier transformation). Note also that the first term in (\ref{final-T}) can also be removed by a constant shift in the $A_i$ field.

The contributions to the longitudinal self-energy in (\ref{final-Long}) can also be traded for ${\cal O}\, (e^4)$ contributions. The redefinition (\ref{fr}) induces the following contribution,
\be
\delta \Pi^{L}_{(1)} (L)= C_0 \, l^2 \, \int \frac{d\Omega}{4\pi} \frac 12 \frac{(v - {\tilde v}) \cdot L}{v \cdot L}=
 C_0 \,l^2 \, \left(1-\frac{l_0}{2l}\lnLL\right) \,.
\ee
Then by taking
\be
C_0=-\frac{\alpha}{3\pi}\left(2-\frac{L^2}{l^2}\right)\,,
\ee
we can remove (\ref{final-Long}). Note again that the operator $C_0$ is non-local in space but local in time.

\end{document}